\documentclass[journal,a4paper]{IEEEtran}
\pdfoutput=1

\usepackage{cite}
\usepackage[pdftex]{graphicx}
\usepackage[cmex10]{amsmath}
\usepackage{url}
\usepackage[caption=false,font=footnotesize]{subfig}

\usepackage[T1]{fontenc}
\usepackage[utf8]{inputenc}

\usepackage{rmsbrt_math}
\usepackage{rmsbrt_IBC}
\usepackage{rmsbrt_algorithm}
\usepackage{rmsbrt_sets}
\usepackage{rmsbrt_colours}

\begin{document}

\title{Joint Coordinated Precoding and Discrete Rate Selection in Multicell MIMO Networks}

\author{Rasmus~Brandt,~\IEEEmembership{Student~Member,~IEEE,}
        and~Mats~Bengtsson,~\IEEEmembership{Senior~Member,~IEEE}%
\thanks{The authors are with the Department of Signal Processing, ACCESS Linn\ae{}us Centre, School of Electrical Engineering, KTH Royal Institute of Technology, Stockholm, Sweden. E-mails: \texttt{rabr5411@kth.se}, \texttt{mats.bengtsson@ee.kth.se}.}}

\maketitle

\begin{abstract}
Many practical wireless communications systems select their transmit rate from a finite set of modulation and coding schemes, which correspond to a set of discrete rates. In this paper, we therefore formulate a joint coordinated precoding and discrete rate selection problem for multiple-input multiple-output (MIMO) multicell networks. Compared to the common assumption of using the continuous Shannon rates as the user utilities, explicitly accounting for the discrete rates more accurately models practical wireless communication systems. The optimization problem that we formulate is combinatorial and non-convex, however, and is thus hard to solve. We therefore rewrite the problem using a discontinuous rate function, which we then bound using its concave envelope in some domain. Based on block coordinate descent, we provide a convergent resource allocation algorithm which can be implemented in a semi-distributed fashion. Numerical performance evaluation shows performance gains when the discrete rates are optimized using our model, as compared to the traditional methods which use the continuous Shannon rates as the user utilities.
\end{abstract}

\begin{figure*}
    \begin{equation} \label{eq:per_stream_SINR}
        \textsf{SINR}_{i_k,n} \left(\uikn, \{ \Vjl \} \right) = \frac{\abssq{\uikn^\herm \Hiki \vikn}}{\sum_{(j,l,m) \neq (i,k,n)} \abssq{\uikn^\herm \Hikj \vjlm} + \sigma_{i_k}^2 \normsq{\uikn}}
    \end{equation}
    \hrule
\end{figure*}

\section{Introduction}
In the literature on multicell multiple-input multiple-output (MIMO) precoding \cite{Gesbert2010}, the user utility is often modelled as the continuous Shannon rate, which describes the rate that can be achieved with vanishingly low error probabilities using long codewords. This is an optimistic model, which further assumes optimal decoders and modulation constellations with infinite granularity. Practical wireless communications systems typically have non of these however. Instead, these systems are often adhering to the bit-interleaved coded modulation (BICM) paradigm \cite[Ch.~7.4.1]{FundamentalsDigitalCommunication}, where the transmit rate is determined by the selection of a channel code and a modulation constellation size. The discrete combinations of codes and constellations are called the modulation and coding schemes (MCSs). Given a signal-to-interference-and-noise ratio (SINR) at the receiver, the highest discrete rate that achieves some acceptable block error rate is then used for the transmissions.

In this work, we consider the case of joint precoder design and discrete rate selection. We model the problem as a system-level optimization problem, where we aim to maximize the weighted sum rate while using minimal amount of power. Since the optimization problem is both combinatorial and non-convex, we first rewrite it using some discontinuous rate functions. These are then bounded by their concave envelopes, in some domain which can be selected by the system designer. After a linearization step, block coordinate descent \cite[Ch.~2.7]{NonlinearProgramming} is applied, resulting in a convergent algorithm which is distributed over the mobile stations. We evaluate our algorithm using numerical simulations. Compared to the state-of-the-art in continuous rate optimization, our algorithm performs well.

Existing work on joint beamforming and discrete rate selection is scarce, and limited to the multiple-input single-output (MISO) and single-input single-output (SISO) models. In \cite{Wai2013}, a convex approximation of the sum rate was proposed for the MISO case. Through a reweighting procedure, some gains over the state-of-the-art in continuous rate optimization was shown. In \cite{Cheng2015}, a mixed integer second order cone program (MISOCP) was formulated for the MISO case. The problem was mathematically reformulated to be applicable to the commercial branch-and-cut solver CPLEX, which numerically gave the optimal solution.  Two heuristics, based on solving a sequence of SOCP problems, were also proposed. In \cite{Wolkerstorfer2013}, the problem was considered for a subcarrier-based SISO system, and an optimal branch-and-bound algorithm was proposed.

Contrary to the previous work, in this paper we consider the problem for the MIMO case. This is an interesting scenario, since it allows for more degrees of freedom in the optimization: both precoders at the transmitters and receive filters at the receivers should be optimized.

\section{System Model} \label{sec:system_model}
We consider a multicell network with $I$ base stations (BSs), collected in the set $\setI = \{ 1, 2, \ldots, I \}$. The $i$th BS serves the mobile stations (MSs) in the set $\setK_i = \{ 1, \ldots, K_i \}$ with data in the downlink. For brevity, we will denote the $k$th MS served by the $i$th BS as $i_k$. The channel between BS $j$ and MS $i_k$ is $\Hikj \in \complexnumbers^{N_{i_k} \times M_i}$. BS $i$ uses a linear precoder $\Vik \in \complexnumbers^{M_i \times d_{i_k}}$ to serve MS $i_k$ with $d_{i_k}$ data streams. At the receiving end, MS $i_k$ applies a linear receive filter $\Uik \in \complexnumbers^{N_{i_k} \times d_{i_k}}$ for interference rejection. The transmitted signal $\xik \in \complexnumbers^{d_{i_k}}$ has zero mean, unit per-stream power, and is i.i.d.\@ over the streams. We denote the $n$th column of $\Vik$ and $\Uik$ as $\vikn$ and $\uikn$, respectively, and assume single-stream decoding in the receivers. With the interfering broadcast channel as the multiuser interaction model, the received filtered signal for the $n$th stream at MS $i_k$ can thus be written as
\begin{align}
    \hat{x}_{i_k,n} &= \uikn^\herm \Hiki \vikn \xikn \label{eq:received_filtered_signal} \\
    &+ \uikn^\herm \hspace{-0.75em} \sum_{\substack{j \in \setI, l \in \setK_j\\m = 1,\ldots,d_{j_l}}} \hspace{-0.5em} \Hikj \vjlm \xjlm + \uikn^\herm \zik, \notag
\end{align}
where $\zik \sim \mathcal{CN} \left( 0, \sigma_{i_k}^2 \matI \right)$ is the thermal noise. The corresponding per-stream SINR is then given by \eqref{eq:per_stream_SINR}, at the top of next page.

The discrete rates that are available to MS $i_k$ are described by the set \mbox{$\setQ_{i_k} = \left\{ q_{i_k}^{(0)}, \ldots, q_{i_k}^{\left( \card{\setQ_{i_k}}-1 \right)} \right\} \subset \realnumbers^+$}, and we assume without loss of generality that \mbox{$0 = q_{i_k}^{(0)} < q_{i_k}^{(1)} < \ldots < q_{i_k}^{\left( \card{\setQ_{i_k}}-1 \right)} < \infty$}. We include the zero rate in order to ensure feasibility in the optimization problem to be formulated. Due to its inclusion, our optimization formulation will also perform implicit user selection. Different MSs may belong to different terminal classes, corresponding to the discrete rates that they can decode, and the sets $\{ \setQ_{i_k} \}_{i \in \setI, k \in \setK_i}$ need thus not be identical. Some examples of discrete rate sets are:
\begin{example}[Discrete rates in WiFi] \label{ex:wifi}
    In the IEEE 802.11ac WiFi standard, code rates between $1/2$ and $5/6$ are combined with constellations ranging from BPSK to 256-QAM \cite{Bejarano2013}. This gives $\setQ = \{ 0, 0.5, 1, 1.5, 2, 3, 4, 4.5, 5, 6, 6.67 \}$ [bits/s/Hz].
\end{example}
\begin{example}[Discrete rates in cellular communication]
    In the 3GPP LTE standard, code rates between $1/8$ to $4/5$ are combined with constellations ranging from QPSK to \mbox{64-QAM} \cite[Sec.~22.4.4.1]{LTEtheUMTSLongTermEvolution}. This gives $\setQ = \{ 0, 0.25, 0.4, 0.5$, $0.67, 1, 1.33, 1.5, 1.6, 2, 2.67, 3, 3.2, 4, 4.5, 4.8 \}$ [bits/s/Hz].
\end{example}

A discrete rate is achievable if the achieved SINR exceeds a pre-determined threshold:
\begin{definition}[Achievable discrete rate] \label{def:achievable_discrete_rate}
    The discrete rate for the $n$th stream of MS $i_k$, $s_{i_k,n} \in \setQ_{i_k}$, is achievable if and only if the SINR for that stream satisfies
    \begin{equation} \label{eq:SINR_constraint}
        \textsf{SINR}_{i_k,n} \left(\uikn, \{ \Vjl \} \right) \geq \beta_{i_k} ( s_{i_k,n} ),
    \end{equation}
    where $\beta_{i_k} : \realnumbers^+ \rightarrow \realnumbers^+$ is a function that maps a discrete rate to its required minimum SINR.
\end{definition}
For a rate $q_{i_k}^{(p)} \in \setQ_{i_k}$, the required SINR $\beta_{i_k}(q_{i_k}^{(p)})$ is typically selected such that the corresponding block error rate (BLER) at the receiver is lower than some $\epsilon_{i_k}^{(p)} > 0$. An example is given by:
\begin{example}[Receiver with constant implementation margin] \label{ex:implementation_margin}
    Given a BLER target of $\epsilon$, assume that the receiver needs a factor $\bar{\beta} \geq 1$ higher SINR than the theoretical minimum.\footnote{This is called the SINR gap approximation \cite[Ch.~9.2.2]{CooperativeCellularWirelessNetworks}.} The discrete rate then satisfies the following Shannon formula \mbox{$s = \log_2 (1 + \beta(s)/\bar{\beta} )$} and the corresponding minimum required SINR is $\beta(s) = \bar{\beta} (2^s - 1)$.
\end{example}

\begin{figure*}
    \centering
    \subfloat[MSE domain, $\eta(e) = e$]{\includegraphics[width=0.33\linewidth]{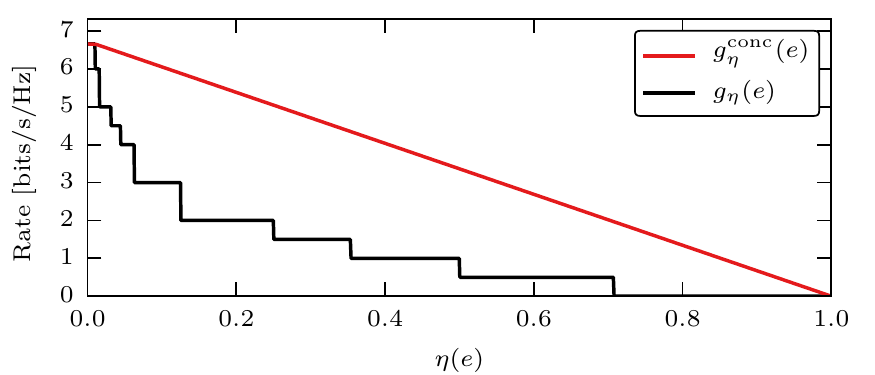} \label{fig:illustrations/approximations-MSE}}
    \subfloat[Continuous rate domain, $\eta(e) = \log_2(e)$]{\includegraphics[width=0.33\linewidth]{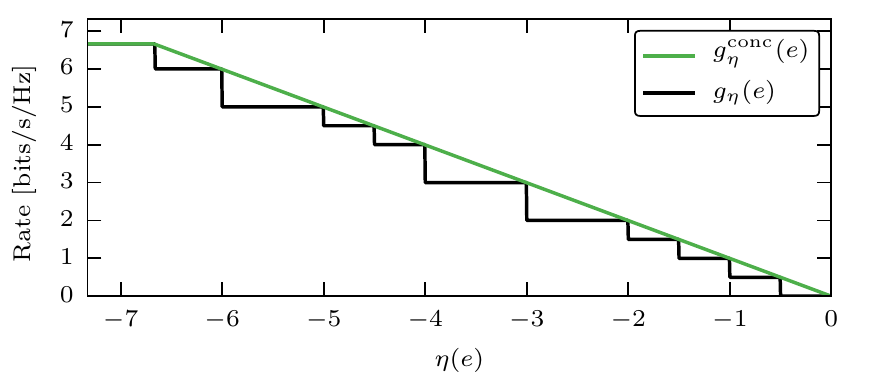} \label{fig:illustrations/approximations-rate}}
    \subfloat[SINR domain, $\eta(e) = 1 - 1/e$]{\includegraphics[width=0.33\linewidth]{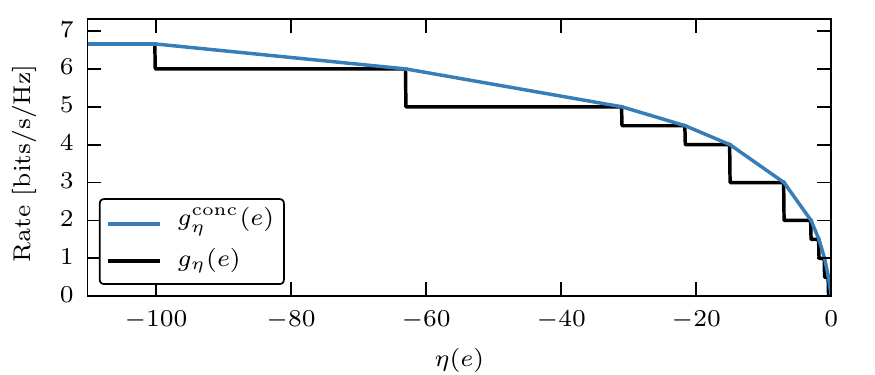} \label{fig:illustrations/approximations-SINR}}
    \caption{Discontinuous rate functions together with their concave envelopes in different QoS domains. Note the poor fit of the bound in (a), compared to the bounds in (b) and (c). The discrete rates were taken from Ex.~\ref{ex:wifi}, with required SINRs from Ex.~\ref{ex:implementation_margin} where $\bar{\beta} = 1$.} \label{fig:illustrations/approximations}
\end{figure*}

\vspace{-2ex}
\section{Joint Coordinated Precoding \\and Discrete Rate Selection}
Our goal is now to optimize the network utility, given the model for the discrete rates. We consider the weighted sum rate as the system-level objective\footnote{For each MS, we sum the discrete rates over all data streams.} function, where $\omega_{i_k} \in \realnumbers_+$ is the weight for MS $i_k$. Since excess power will increase the interference in the network, we maximize the weighted sum rate subject to a power regularization term:
\begin{align}
    & \underset{\substack{\{ \Uik \}, \{ \Vik \}, \\ \{ s_{i_k,n} \}}}{\text{maximize}} & & \sum_{\substack{i \in \setI, k \in \setK_i\\n = 1,\ldots,d_{i_k}}} \hspace{-0.5em} \omega_{i_k} s_{i_k,n} - g_\kappa(\{ \Vjl \}) \notag \\
    & \text{subject to} & & s_{i_k,n} \in \setQ_{i_k}, \quad \forall \, i_k, n \notag \\
    & & & \textsf{SINR}_{i_k,n} \left(\uikn, \{ \Vjl \} \right) \geq \beta_{i_k} (s_{i_k,n}), \; \forall \, i_k, n \notag \\  
    & & & \sum_{k \in \setK_i} \Fnormsq{\Vik} \leq P_i, \quad \, \forall \, i \in \setI, \label{opt:original}
\end{align}
where $g_\kappa(\{ \Vjl \}) = \kappa \sum_{i \in \setI, k \in \setK_i} \Fnormsq{\Vik}$. The regularization parameter $\kappa$ is selected according to the following lemma, as inspired by Claim~1 in \cite{Matskani2008}:
\begin{lemma} \label{lemma:power_regularization}
    Define $f(\{ s_{i_k,n} \}) = \sum_{\substack{i \in \setI, k \in \setK_i\\n = 1,\ldots,d_{i_k}}} \omega_{i_k} s_{i_k,n}$ and $\delta~=~\min_{\{ s_{i_k,n} \}, \{ \check{s}_{i_k} \}, \{ s_{i_k,n} \} \neq \{ \check{s}_{i_k,n} \}} \abs{f(\{ s_{i_k,n} \}) - f(\{ \check{s}_{i_k,n} \})}$. If $\kappa = \frac{\delta}{\sum_{i \in \setI} P_i + 1}$, the solution to the optimization problem in \eqref{opt:original} simultaneously gives the maximum weighted sum rate and the corresponding minimum sum power precoders.
\end{lemma}
\begin{IEEEproof}
    Similar to the proof of \cite[Claim~1]{Matskani2008}.
\end{IEEEproof}
With this selection of $\kappa$, no loss in the objective due to selecting a smaller discrete rate for some MS can be made up for by the corresponding decrease in used power. Therefore, the optimization problem in \eqref{opt:original} simultaneously gives the maximum weighted sum rate and the corresponding minimum sum power precoders. This hinges on the facts that the weighted sum rate only takes on discrete values and that the sum power is bounded; see related discussion in \cite{Matskani2008}.

\vspace{-3ex}
\subsection{Problem Reformulation}
The optimization problem in \eqref{opt:original} is both combinatorial (due to the selection of the discrete rates) and non-concave (due to the non-concavity of $\textsf{SINR}_{i_k,n} \left(\uikn, \{ \Vjl \} \right)$). As posed, it is thus difficult to solve. We will therefore reformulate the problem into one with a discontinuous objective function, which we will then bound. After the reformulation and bounding, we will apply the ideas pioneered in \cite{Christensen2008,Shi2011} for the optimization. This entails linearizing the objective and applying block coordinate descent \cite[Ch.~2.7]{NonlinearProgramming}.

The first step in the reformulation is the introduction of the mean squared error (MSE) of the $n$th stream of MS $i_k$:
\begin{align}
    e_{i_k,n}&(\uikn, \{ \Vjl \}) = \ex{\abssq{x_{i_k,n} - \hat{x}_{i_k,n}}} = \\
    &1 - 2\real{\uikn^\herm \Hiki \vikn} + \uikn^\herm \Phiik \uikn, \notag
\end{align}
where the received signal covariance matrix for MS $i_k$ is \mbox{$\Phiik = \sum_{j \in \setI, l \in \setK_j} \Hikj \Vjl \Vjl^\herm \Hikj^\herm + \sigma_{i_k}^2 \matI$}. Assuming finite-power precoders, together with the unit-power symbols, we have that $ 0 < e_{i_k,n}(\uikn, \{ \Vjl \}) \leq 1, \, \forall \, i_k, n$.

Next, we rewrite the SINR constraint as a general quality of service (QoS) constraint, which is a function of the MSE.
\begin{definition}[QoS domain] \label{def:QoS_domain}
    Let $\eta \! : \realnumbers_+ \rightarrow \realnumbers_+$ be a concave and strictly increasing function. It describes the mapping from the MSE domain to another \emph{QoS domain}.
\end{definition}
\begin{definition}[Discontinuous rate function] \label{def:discontinuous_rate_function}
    Given a fixed receive filter $\uikn$, fixed precoders $\{ \Vjl \}_{j \in \setI, l \in \setK_j}$, and a QoS domain represented by $\eta(\cdot)$, the discrete rate for the $n$th stream of MS $i_k$ is given by the \emph{discontinuous rate function}
    \begin{align}
        g_\eta (&e_{i_k,n}(\uikn, \{ \Vjl \})) = \label{eq:discontinuous_rate_function} \\
        &\begin{aligned}
            & \underset{q \in \setQ_{i_k}}{\textnormal{maximize}} & & q \notag \\
            & \textnormal{subject to} & & \eta \left( e_{i_k,n}(\uikn, \{ \Vjl \}) \right) \leq \eta \left( \frac{1}{1+\beta_{i_k}(q)} \right). \notag
        \end{aligned}
    \end{align}
\end{definition}
In the following, this discontinuous function will be bounded by a continuous function. By introducing $\eta(\cdot)$ into \eqref{eq:discontinuous_rate_function}, we get a degree of freedom in designing this bound.

Given Def.~\ref{def:discontinuous_rate_function}, we now reformulate the problem in \eqref{opt:original} as:
\begin{align}
    & \underset{\{ \Uik \}, \{ \Vik \}}{\text{maximize}} & & \hspace{-3mm} \sum_{\substack{i \in \setI, k \in \setK_i\\n = 1,\ldots,d_{i_k}}} \hspace{-3mm} \omega_{i_k} g_\eta(e_{i_k,n}(\uikn, \{ \Vjl \})) \notag - g_\kappa(\{ \Vjl \}) \notag \\
    & \text{subject to} & & \sum_{k \in \setK_i} \Fnormsq{\Vik} \leq P_i, \quad \, \forall \, i \in \setI. \label{opt:transformed}
\end{align}
The discrete rates are now implicitly selected in \eqref{eq:discontinuous_rate_function}, and the problem is no longer combinatorial. The objective function has however become discontinuous. There is no loss in optimality due to this reformulation though, since it holds that
\begin{equation*}
    \min_{\uikn} e_{i_k,n}(\uikn, \{ \Vjl \}) = \min_{\uikn} \frac{1}{1 + \textsf{SINR}_{i_k,n} \left(\uikn, \{ \Vjl \} \right)}.
\end{equation*}

We will now bound the objective function in \eqref{opt:transformed} by bounding the discontinuous rate function by its concave envelope.\footnote{The concave envelope is the ``smallest'' concave function which majorizes the function. It is thus the best concave approximation available.} Given $\setP_{i_k} \subset \naturalnumbers$ and $\{ c_{i_k}^{(p)} \}_{p \in \setP_{i_k}}$, $\{ m_{i_k}^{(p)} \}_{p \in \setP_{i_k}}$ which are uniquely defined slopes and offsets,\footnote{These are uniquely determined by $\setQ_{i_k}$ and $\beta_{i_k}(\cdot)$, see examples in Fig.~\ref{fig:illustrations/approximations}.} the concave envelope is given by the following piecewise linear function:
\begin{equation}
    g_\eta^\text{conc} (e) = \underset{p \in \setP_{i_k}}{\text{min}} \left\{ c_{i_k}^{(p)} \eta \left( e \right) + m_{i_k}^{(p)} \right\} \geq g_\eta(e), \notag
\end{equation}
Some examples of concave envelopes are given at the top of the page, in Fig.~\ref{fig:illustrations/approximations}, for three different QoS domains. This figure illustrates two key properties of our model. First, note that different QoS domains give bounds with different tightness. In the continuous rate domain (i.e. $\eta(e) = \log_2(e)$) for example, the concave envelope is a tight bound. This is because the discrete rates ``look linear'' in the this domain (cf. Ex.~\ref{ex:implementation_margin}), and are thus well approximated by a piecewise linear function. The second property to note is that our model accounts for the maximum discrete rate that is achievable. There is thus no point in reducing the MSE past the threshold value of the largest discrete rate.\footnote{By reducing the MSE further, the performance at the corresponding MS would not increase but all other MSs might receive stronger interference, which is detrimental to the system-level performance.} In Figure~\ref{fig:illustrations/approximations}, this is seen by the curves having zero slope for sufficiently small MSEs.

By bounding the discontinuous rate functions by their concave envelopes, we get the following optimization problem:
\begin{align}
    & \underset{\{ \Uik \}, \{ \Vik \}}{\text{maximize}} & & \hspace{-3mm} \sum_{\substack{i \in \setI, k \in \setK_i\\n = 1,\ldots,d_{i_k}}} \hspace{-3mm} \omega_{i_k} g_\eta^\text{conc}(e_{i_k,n}(\uikn, \{ \Vjl \})) - g_\kappa(\{ \Vjl \}) \notag \\
    & \text{subject to} & & \sum_{k \in \setK_i} \Fnormsq{\Vik} \leq P_i, \quad \, \forall \, i \in \setI. \label{opt:conc}
\end{align}
This step introduces some non-optimality, since we are upper bounding the objective of a maximization problem. The problem is no longer discontinuous however. The final hurdle is now the non-concavity of the objective. By Taylor expanding the $\eta(\cdot)$ function around a point $1/\wikn$, we get
\begin{align*}
    &g_{\eta^\text{lin}}^\text{conc}(e_{i_k,n}(\uikn, \{ \Vjl \}), \wikn) = \\
    &\underset{p \in \setP_{i_k}}{\text{min}} \Bigg\{ c_{i_k}^{(p)} \Bigg( \eta \left( \frac{1}{\wikn} \right) + \eta' \left( \frac{1}{\wikn} \right) \Bigg( e_{i_k,n}(\uikn, \{ \Vjl \}) \\
    &\hspace{10mm}- \frac{1}{\wikn} \Bigg) \Bigg) + m_{i_k}^{(p)} \Bigg\} \leq g_{\eta}^\text{conc}(e_{i_k,n}(\uikn, \{ \Vjl \})).
\end{align*}
The inequality holds since $c_{i_k}^{(p)} \eta(\cdot)$ is a convex function\footnote{By construction, it holds that $c_{i_k}^{(p)} \leq 0, \, \forall \, p \in \setP_{i_k}$.} together with the fact that the first-order Taylor expansion of a convex function is a global underestimator \cite[Ch.~3.1.3]{ConvexOptimization}. By introducing the linearization points as optimization variables, we get the final optimization problem as:
\begin{align}
    & \underset{\substack{\{ \Uik \}, \{ \Vik \}, \\ \{ \wikn \}}}{\text{maximize}} & & \hspace{-3mm} \sum_{\substack{i \in \setI, k \in \setK_i\\n = 1,\ldots,d_{i_k}}} \hspace{-3mm} \omega_{i_k} g_{\eta^\text{lin}}^\text{conc} (e_{i_k,n}, \wikn) - g_\kappa(\{ \Vjl \}) \notag \\
    & \text{subject to} & & \sum_{k \in \setK_i} \Fnormsq{\Vik} \leq P_i, \quad \, \forall \, i \in \setI. \label{opt:conc_lin}
\end{align}
It can easily be shown that $\max_{\wikn} g_{\eta^\text{lin}}^\text{conc}(\cdot, \wikn) = g_\eta^\text{conc}(\cdot)$, i.e., the linearization is tight at optimality. The optimization problems in \eqref{opt:conc} and \eqref{opt:conc_lin} therefore have the same optimal value.

\subsection{Distributed Algorithm}
The final optimization problem in \eqref{opt:conc_lin} has the desired property that it is concave in each block of variables, when the two other blocks are held fixed. This leads us to apply block coordinate descent (BCD) \cite[Ch.~2.7]{NonlinearProgramming} to it.

By fixing the precoders and linearization weights in the optimization problem in \eqref{opt:conc_lin}, it can be shown that an optimal receive filter is the MMSE filter \mbox{$\Uik^\star = \Phiik^{-1} \Hiki \Vik, \, \forall \, i_k$}. By fixing the receive filters and the precoders, it can be shown that optimal linearization weights are \mbox{$\wikn^\star = 1/e_{i_k,n}(\uikn^\star, \{ \Vjl \}), \, \forall \, i_k, n$}. Finally, the optimal precoders are given by the optimization problem when the receive filters and linearization weights are fixed. This strongly concave optimization problem has a unique solution, which can be found using, e.g., interior-point methods \cite[Ch.~11]{ConvexOptimization}.

By sequentially solving the subproblems, an iterative algorithm is obtained. The receive filters and linearization weights can be solved for distributedly over the MSs, whereas the precoders must be solved for centrally at the BSs.
\begin{theorem}
    When BCD is applied to the optimization problem in \eqref{opt:conc_lin}, the sequence of objective values obtained converges.
\end{theorem}
\begin{IEEEproof}
    The sequence of objective values is nondecreasing, since in each step of the BCD, the objective function is maximized. The sequence is further bounded above by the finite optimal value of the optimization problem in \eqref{opt:conc_lin}. The sequence thus converges \cite[Thm.~3.14]{PrinciplesofMathematicalAnalysis}.
\end{IEEEproof}

\begin{figure}[t]
    \centering
    \includegraphics{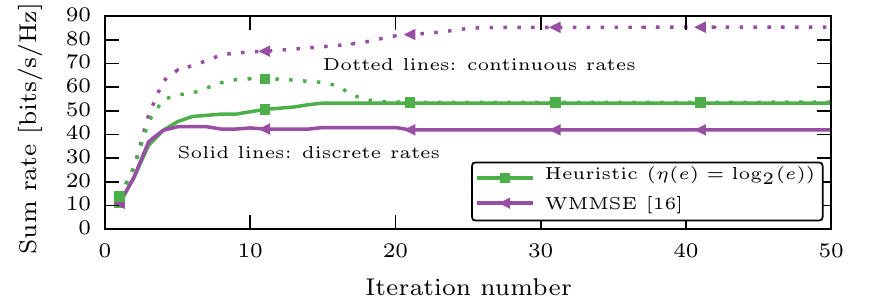}
    \vspace{-5ex}
    \caption{Example of convergence of the algorithms for one realization.} \label{fig:convergence}
    \vspace{1.5ex}
    \includegraphics{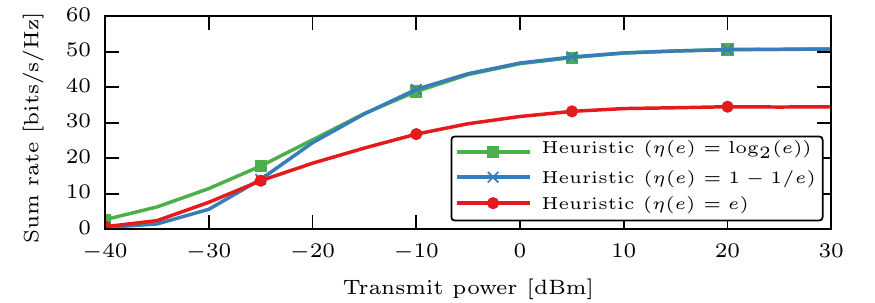}
    \vspace{-5ex}
    \caption{Comparing different QoS domains.} \label{fig:SNR-heuristics}
    \vspace{1.5ex}
    \includegraphics{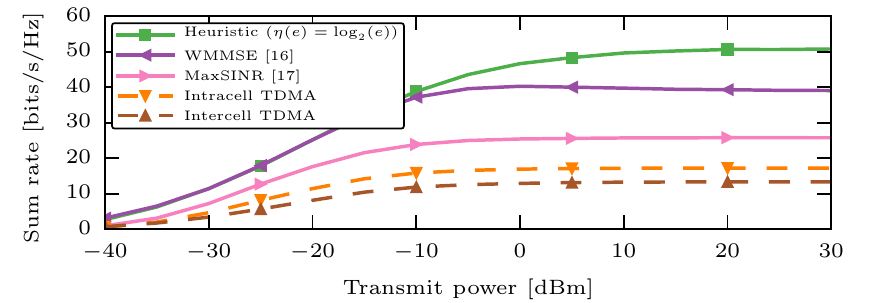}
    \vspace{-5ex}
    \caption{Comparing the proposed algorithm to the benchmarks.} \label{fig:SNR-baselines}
    \vspace{1.5ex}
    \includegraphics{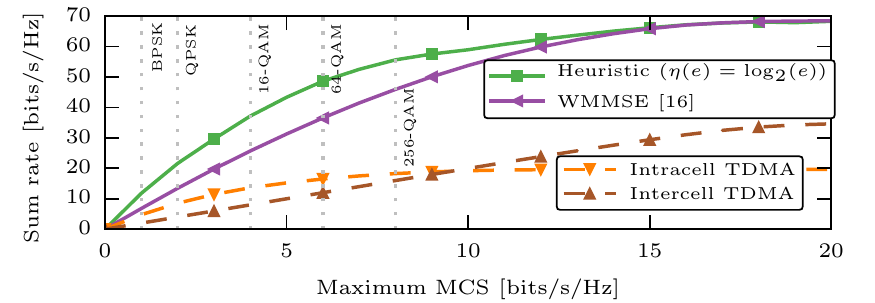}
    \vspace{-5ex}
    \caption{Sum rate when varying the maximum discrete rate. The rates of uncoded QAM-constellations are shown in grey dotted lines.} \label{fig:Q}
    \vspace{-2ex}
\end{figure}

\vspace{-2ex}
\section{Performance Evaluation}
We evaluate the performance of the proposed algorithm using numerical simulations. We let $I = 3$ BSs be placed equidistant along the centre line of a $120 \times 20$ [m] office corridor. Each BS serves $K = 2$ randomly placed MSs with $d = 2$ data streams. The BSs have $M = 4$ antennas each and the MSs have $N = 2$ antennas each. The large-scale fading is given by the ITU-R InH model \cite[Table~A1-2]{ITURInH}, but we model the small-scale fading as i.i.d.\@ Rayleigh fading. We use the discrete rates from Ex.~\ref{ex:wifi} with corresponding required SINRs from Ex.~\ref{ex:implementation_margin} with $\bar{\beta} = 1$. We draw $100$ i.i.d.\@ Monte Carlo realizations, and average the results. The proposed algorithm is run until the relative difference between subsequently achieved objective values is less than $10^{-3}$. We compare our proposed algorithm to the per-stream WMMSE algorithm \cite{Komulainen2013} and the MaxSINR algoritm \cite{Gomadam2011}, which both are well-known to perform well for the continuous rate\footnote{We define the continuous rate as  $\log_2(1 + \textsf{SINR})$.} case \cite{Schmidt2013}.\footnote{The existing work in \cite{Wolkerstorfer2013,Wai2013,Cheng2015} cannot handle the MIMO case, which we consider here, and are consequently not included as benchmarks.} We also consider intercell and intracell time-division multiple access (TDMA), where the precoders are given by waterfilling over the strongest singular vectors of the desired channel.

In Fig.~\ref{fig:convergence}, we show the convergence of our algorithm when $\eta(e) = \log(e)$ and the transmit power is 21 dBm. The achieved discrete and continuous rates are more interesting than the (regularized) optimization objectives, and we thus show the former. After about 20 iterations, the two rates converge. This indicates that no excess power is used, since otherwise the continuous rate would be larger than the discrete rate. The discrete rate performance of the WMMSE algorithm is poor since it allocates too much power to already saturated streams.

In Fig.~\ref{fig:SNR-heuristics}, we compare the performance for different QoS domains. The rate and SINR domains perform identically at high transmit powers, whereas the MSE domain is unable to perform as well (cf. the bound tightness in Fig.~\ref{fig:illustrations/approximations}). In Fig.~\ref{fig:SNR-baselines}, we compare our algorithm to the benchmarks. At high transmit powers, the necessity of modelling the discrete rates is clear. The WMMSE algorithm performs worse for sufficiently high transmit power and the other benchmarks are not competitive.

In Fig.~\ref{fig:Q}, we vary the number of available discrete rates at a fixed transmit power of 21 dBm. We consider discrete rates $\setQ = \{ 1, 2, \ldots, q^\text{max} \}$, where we sweep $q^\text{max}$ in steps of 1 [bits/s/Hz]. At typical constellation sizes, our proposed algorithm is clearly superior. For very large constellations however, the WMMSE algorithm catches up.

\vspace{-2ex}
\section{Conclusion}
Many practical wireless communications systems use a finite set of discrete rates. By explicitly modelling these, a heuristic coordinated precoding algorithm was developed, which performs very well compared to algorithms which do not account for the discrete rates.

\bibliographystyle{IEEEtran}
\bibliography{IEEEabrv,/Users/rasmus/phd/texmf/bibtex/bib/coordinated_precoding,/Users/rasmus/phd/texmf/bibtex/bib/rasmus_brandt}

\begin{thebibliography}{10}
\providecommand{\url}[1]{#1}
\csname url@samestyle\endcsname
\providecommand{\newblock}{\relax}
\providecommand{\bibinfo}[2]{#2}
\providecommand{\BIBentrySTDinterwordspacing}{\spaceskip=0pt\relax}
\providecommand{\BIBentryALTinterwordstretchfactor}{4}
\providecommand{\BIBentryALTinterwordspacing}{\spaceskip=\fontdimen2\font plus
\BIBentryALTinterwordstretchfactor\fontdimen3\font minus
  \fontdimen4\font\relax}
\providecommand{\BIBforeignlanguage}[2]{{%
\expandafter\ifx\csname l@#1\endcsname\relax
\typeout{** WARNING: IEEEtran.bst: No hyphenation pattern has been}%
\typeout{** loaded for the language `#1'. Using the pattern for}%
\typeout{** the default language instead.}%
\else
\language=\csname l@#1\endcsname
\fi
#2}}
\providecommand{\BIBdecl}{\relax}
\BIBdecl

\bibitem{Gomadam2008}
K.~Gomadam, V.~R. Cadambe, and S.~A. Jafar, ``Approaching the capacity of
  wireless networks through distributed interference alignment,'' in
  \emph{Proc. IEEE Global Telecommun. Conf. (GLOBECOM'08)}, 2008, pp. 1--6.

\bibitem{Gesbert2010}
D.~Gesbert, S.~Hanly, H.~Huang, S.~Shamai~Shitz, O.~Simeone, and W.~Yu,
  ``Multi-cell {MIMO} cooperative networks: A new look at interference,''
  \emph{{IEEE} J. Sel. Areas Commun.}, vol.~28, no.~9, pp. 1380--1408, 2010.

\bibitem{FundamentalsDigitalCommunication}
U.~Madhow, \emph{Fundamentals of Digital Communication}.\hskip 1em plus 0.5em
  minus 0.4em\relax Cambridge University Press, 2008.

\bibitem{NonlinearProgramming}
D.~Bertsekas, \emph{Nonlinear programming}.\hskip 1em plus 0.5em minus
  0.4em\relax Athena Scientific, 2006.

\bibitem{Wai2013}
H.-T. Wai, Q.~Li, and W.-K. Ma, ``A convex approximation method for multiuser
  {MISO} sum rate maximization under discrete rate constraints,'' in
  \emph{Proc. IEEE Int. Conf. Acoustics, Speech, Signal Process. (ICASSP'13)},
  May 2013, pp. 4759--4763.

\bibitem{Cheng2015}
Y.~Cheng and M.~Pesavento, ``Joint discrete rate adaptation and downlink
  beamforming using mixed integer conic programming,'' \emph{{IEEE} Trans.
  Signal Process.}, vol.~63, no.~7, pp. 1750--1764, Jan. 2015.

\bibitem{Wolkerstorfer2013}
M.~Wolkerstorfer, J.~Jald{\'e}n, and T.~Nordstr{\"o}m, ``Low-complexity optimal
  discrete-rate spectrum balancing in digital subscriber lines,'' \emph{Signal
  Processing}, vol.~93, no.~1, pp. 23--34, Jan. 2013.

\bibitem{Bejarano2013}
O.~Bejarano, E.~W. Knightly, and M.~Park, ``{IEEE} 802.11ac: From
  channelization to multi-user {MIMO},'' \emph{{IEEE} Commun. Mag.}, 2013.

\bibitem{LTEtheUMTSLongTermEvolution}
S.~Sesia, I.~Toufik, and M.~Baker, \emph{{LTE}: the {UMTS} long term
  evolution}.\hskip 1em plus 0.5em minus 0.4em\relax Wiley, 2009.

\bibitem{CooperativeCellularWirelessNetworks}
E.~Hossain, D.~I. Kim, and V.~K. Bhargava, Eds., \emph{Cooperative Cellular
  Wireless Networks}.\hskip 1em plus 0.5em minus 0.4em\relax Cambridge
  University Press, 2011.

\bibitem{Matskani2008}
E.~Matskani, N.~Sidiropoulos, Z.-Q. Luo, and L.~Tassiulas, ``Convex
  approximation techniques for joint multiuser downlink beamforming and
  admission control,'' \emph{{IEEE} Trans. Wireless Commun.}, vol.~7, no.~7,
  pp. 2682--2693, Jul. 2008.

\bibitem{Christensen2008}
S.~Christensen, R.~Agarwal, E.~Carvalho, and J.~Cioffi, ``Weighted sum-rate
  maximization using weighted {MMSE} for {MIMO-BC} beamforming design,''
  \emph{{IEEE} Trans. Wireless Commun.}, vol.~7, no.~12, pp. 4792--4799, Dec.
  2008.

\bibitem{Shi2011}
Q.~Shi, M.~Razaviyayn, Z.-Q. Luo, and C.~He, ``An iteratively weighted {MMSE}
  approach to distributed sum-utility maximization for a {MIMO} interfering
  broadcast channel,'' \emph{{IEEE} Trans. Signal Process.}, vol.~59, no.~9,
  pp. 4331--4340, 2011.

\bibitem{ConvexOptimization}
S.~Boyd and L.~Vandenberghe, \emph{Convex Optimization}.\hskip 1em plus 0.5em
  minus 0.4em\relax Cambridge University Press, 2004.

\bibitem{PrinciplesofMathematicalAnalysis}
W.~Rudin, \emph{Principles of Mathematical Analysis}, 3rd~ed.\hskip 1em plus
  0.5em minus 0.4em\relax McGraw-Hill, 1976.

\bibitem{ITURInH}
ITU-R, ``Guidelines for evaluation of radio interface technologies for
  {IMT-Advanced},'' ITU-R, Tech. Rep. M.2135-1, 2009.

\bibitem{Komulainen2013}
P.~Komulainen, A.~T\"{o}lli, and M.~Juntti, ``Effective {CSI} signaling and
  decentralized beam coordination in {TDD} multi-cell {MIMO} systems,''
  \emph{{IEEE} Trans. Signal Process.}, vol.~61, no.~9, pp. 2204--2218, May
  2013.

\bibitem{Gomadam2011}
K.~Gomadam, V.~R. Cadambe, and S.~Jafar, ``A distributed numerical approach to
  interference alignment and applications to wireless intererence networks,''
  \emph{{IEEE} Trans. Inf. Theory}, vol.~57, no.~6, pp. 3309--3322, 2011.

\bibitem{Schmidt2013}
D.~Schmidt, C.~Shi, R.~Berry, M.~Honig, and W.~Utschick, ``Comparison of
  distributed beamforming algorithms for {MIMO} interference networks,''
  \emph{{IEEE} Trans. Signal Process.}, vol.~61, no.~13, pp. 3476--3489, 2013.

\end{thebibliography}

\end{document}